\documentclass[a4paper,fleqn,usenatbib]{mnras}

\usepackage{newtxtext,newtxmath}

\usepackage[T1]{fontenc}
\usepackage{ae,aecompl}

\usepackage{graphicx}   
\usepackage{amsmath}    
\usepackage{amssymb}    
\usepackage{makecell}
\usepackage{threeparttable}

\title[Photoionization-driven NAL variability]{Evidence for photoionization-driven variability in narrow absorption lines}

\author[W.-J. Lu et al.]{
Wei-Jian Lu,$^{1}$\footnotemark[1]
Ying-Ru Lin,$^{1}$\footnotemark[1]
Yi-Ping Qin,$^{1,2}$Wei-Rong Huang,$^{2}$Cai-Juan Pan,$^{1}$
\newauthor
{ }Hong-Yan Huang,$^{1}$Min Yao,$^{1}$Wei-Jing Nong,$^{1}$Zhi-Kao Yao,$^{1}$Mei-Mei Lu,$^{1}$
\newauthor
{ }Da-Sheng Pan,$^{1}$Hong-Qiang Huang,$^{1}$and Qing-Lin Han$^{1}$
\\
$^{1}$Dapartment of Physics and Telecommunication Engineering, Baise University, Baise 533000, China\\
$^{2}$Center for Astrophysics, Guangzhou University, Guangzhou 510006, China}

\date{Accepted XXX. Received YYY; in original form ZZZ}

\pubyear{2017}

\begin{document}
\label{firstpage}
\pagerange{\pageref{firstpage}-\pageref{lastpage}}
\maketitle
\begin{abstract}
In this Letter, we report the discovery of a strong correlation
between the variability of narrow absorption lines (NALs) and the
ionizing continuum  from a two-epoch spectra sample of 40 quasars
containing 52 variable \ion{C}{iv} $\mathrm{\lambda\lambda}1548,
1551$ absorption doublets. According to the concordance index, this
sample is classified into two subsamples. Subsample I shows an
anti-correlation between the variations of absorption lines and the
continuum, while Subsample II exhibits a positive correlation. These results
imply that these variable \ion{C}{iv} $\lambda\lambda1548, 1551$
absorption doublets are intrinsic to the corresponding quasars and
that their variations are caused primarily by the fluctuations of the
ionizing continuum. Based on our analysis, we propose that there might
be two kinds of absorption gas: one that is very sensitive to the
continuum variations, the another that is not. In addition, we suggest that in many cases the emergence or disappearance of NALs is caused by fluctuations of the ionizing continuum.

\end{abstract}


\footnotetext[1]{E-mail:william\_lo@qq.com; yingru\_lin@qq.com}


\section{Introduction}
The absorption lines in quasar spectra can place constraints on the
basic properties of quasar environments and be used to probe the gaseous
content of the Universe from early cosmic times to the present day.
According to their origins, quasar absorption lines are divided into two classes. Intrinsic absorption lines are thought to
be produced by the gas directly associated with the quasar central
region or by that in the host galaxy, while intervening absorption
lines are caused by foreground galaxies along the line of sight to
the background quasar. Based on the absorption width of
their profiles, intrinsic absorption lines can be further classified
into broad absorption lines (BALs: absorption widths of at
least 2000 $\rm km~s^{-1}$, e.g. Weymann et al. 1991), narrow
absorption lines (NALs: absorption widths of a few hundred
$\rm km~s^{-1}$) and mini-broad absorption lines (mini-BALs:
absorption widths between BALs and NALs; e.g. Hamann \&
Sabra 2004).

Previous observations have found variations (both in strength and in 
shape) in quasar BALs/mini-BALs on time-scales ranging from a few years to a
few months in the quasar rest-frame (e.g. Filiz Ak et al. 2012,
2013; Welling et al. 2014; Misawa, Charlton \& Eracleous 2014; Wang
et al. 2015, and references therein), and similar variations in NALs (e.g.
Hamann et al. 1995; Barlow, Hamann \& Sargent 1997; Hamann,
Barlow \& Junkkarinen 1997; Ganguly, Charlton \& Eracleous 2001b;
Narayanan et al. 2004; Wise et al. 2004; Misawa et al.
2005; Hamann et al. 2011; Chen et al. 2013, 2015; Chen \& Qin
2015). The time variability of absorption lines may be caused by the
bulk motion of the absorbing gas across the line of sight
(e.g., Hamann et al. 2008; Krongold, Binette \&
Hern$\'a$ndez-Ibarra 2010; Hall et al. 2011; Capellupo et al. 2013;
Chen et al. 2013; Shi et al.
2016; Rogerson et al. 2016) or by changes in the ionization of
the gas (e.g., Hamann et al. 2011; Filiz Ak et al. 2013; Arav et al. 2015; Chen \& Qin 2015; Wang et al.
2015). Which mechanism is the main cause of the absorption-line
variability is still under debate.

Research into the correlation between variations in absorption lines and the
continuum is a method that has been proposed for determining the origin of the
absorption-line variability. In fact, some researches into the correlations between
variations of BALs and the continuum/emission has already been carried out. A few papers reported that no clear correlation between
the variation of BALs and the continuum (Gibson et al. 2008; Wildy
et al. 2014; Vivek et al. 2014) has ever been detected, while
several recent studies have provided evidence of the coordinated variations
between BALs and the continuum (Capellupo et al. 2012; Filiz Ak et
al. 2012, 2013; Misawa et al. 2014; Wang et al. 2015).
Although there are no reports on the correlation between the
variability of NALs and the continuum, recent researches based on
combining multi-epoch high-resolution UV and X-ray data for the Seyfert
galaxy NGC 5548 showed a definitive case of NAL variation caused by
photoionization changes (Kaastra et al. 2014, Arav et al. 2015),
and well-coordinated variations of different absorption systems
between different epochs have been detected in individual quasars
(Hamann et al. 2011; Chen \& Qin 2015).

In this paper, we study the correlation between the variations of C
{\footnotesize IV} NALs and those of the continuum. We describe the sample  of spectra in Section 2 and perform statistical analyses in Section 3. The discussion is presented in Section 4, and conclusions are
provided in Section 5.

\section{Spectral sample}
The Sloan Digital Sky Survey (SDSS; York et al. 2000) is a project
that aims to obtain detailed three-dimensional maps of a large area of the
Universe using a 2.5-m telescope (Gunn et al. 2006). The
first three periods of the survey in the SDSS project have been completed.
The Baryon Oscillation Spectroscopic Survey (BOSS; Eisenstein et al.
2011) is one of the programs in the third period of SDSS
(SDSS-III, Eisenstein et al. 2011; $\rm P\hat{a}ris$ et al.
2012). SDSS-I/II spectra have a spectral resolution of
\emph R$\approx$1800-2200 (e.g. York et al. 2000), while BOSS spectra have
a resolution of \emph R$\approx$ 1300-2500 (P$\hat{\rm a}$ris et al.
2012).

Using the sample of 7932 quasars observed by both SDSS-I/II (Data
Release 7; Schneider et al. 2010) and BOSS (Data Release 9; $\rm
P\hat{a}ris$ et al. 2012), Chen et al. (2015) presented a catalogue
of 52 pairs of obviously variable \ion{C}{iv} $\lambda\lambda1548,
1551$ absorption doublets (with confident levels of $\textgreater$4$\sigma$ for
$\lambda$ 1548 lines and $\textgreater$3$\sigma$ for $\lambda$ 1551 lines),
identified from 40 quasar spectra. In the catalogue of
variable \ion{C}{iv} absorption doublets, there are 24 systems that have
emerged in or disappeared from the latter spectra. The two SDSS
observations span timescales $\Delta \rm{MJD}$ = 304--1416 d in the
quasar rest-frame. The range of the emission redshift of these
quasars is $z_{\rm em}$ = 1.8--3.5, while that of the absorption
redshift of the 52 \ion{C}{iv} $\lambda\lambda1548, 1551$
absorption doublets is $z_{\rm abs}$ = 1.7--3.3. All of the
40 quasars have a signal-to-noise ratio (SNR) $\geq$8.0 for
both SDSS-I/II and BOSS spectra (Chen et al. 2015).

In the following, we will study the correlation between the
variations of NALs and those of the continuum using the above sample
consisting of the two-epoch spectra of the 40 quasars.

\section{Correlation analysis and results}
\subsection{Distribution of the concordance index for C {\small IV} NALs against the continuum}
\begin{figure}
\includegraphics[width=\columnwidth]{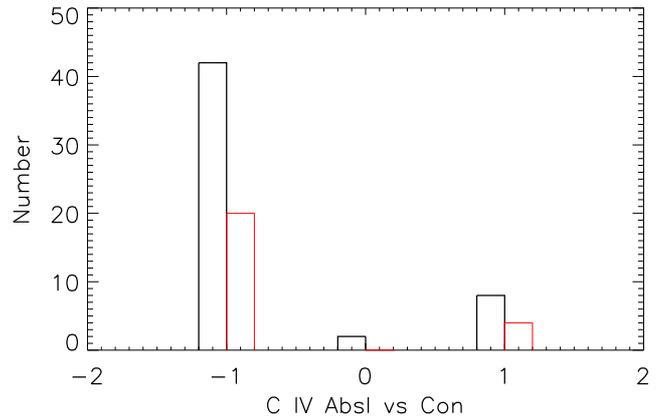}
 \caption{Count distribution of the concordance index for C {\scriptsize IV} NALs against the continuum. A concordance index of $+1$ represents the cases where both the absorption line and the continuum become stronger or both of them become weaker, one of $-1$ represents the opposite cases, and one of $0$ represents the case in which the continuum does not change significantly between the two SDSS observations. The black boxes show the distribution of all spectrum pairs in our sample, while the red boxes show the distribution of the spectrum pairs of the  emergence or disappearance of C {\scriptsize IV} NALs.}
    \label{fig:1}
\end{figure}

Adopting the method used by Wang et al. (2015), we consider that the
continuum varies significantly when the fluctuation of the
continuum flux at $\thicksim$1450 {\rm \AA} (in the quasar rest-frame) is
more than 5 per cent between the two observations. In this section,
the variations of the equivalent widths (EWs) of C {\small
IV} absorption lines are taken directly from Chen et al. (2015).
According to Wang et al. (2015), we define a concordance index of
$+1$ for the cases in which both the absorption line and the continuum
become stronger or both of them become weaker, of $-1$
when they vary in the opposite way, and
of $0$ when the continuum has no significant
variation. The distribution of the concordance index is shown in
Fig.~\ref{fig:1}.

As depicted in Fig.~\ref{fig:1}, the variations of absorption-line EWs are
highly statistically coordinated with the variations of the
continuum flux. More specifically, about 81 per cent of the \ion{C}{iv} absorption lines weaken when the continuum brightens, or
the absorption lines strengthen while the continuum dims (the case
when the concordance index is $-1$); about 15 per cent of the \ion{C}{iv} absorption lines strengthen when the continuum brightens,
or the absorption lines weaken while the continuum dims (the case when
the concordance index is $+1$); the remaining two quasars have no
significant continuum variations between the two epochs of
observations. Note that a highly level of coordination is also detected from
the spectrum pairs of the emerged or disappeared events of C {\small
IV} NALs.
\begin{table*}
    \centering
    \caption{C {\scriptsize IV} absorption systems and the continuum data.}
    \label{tb:1}
    \begin{tabular}{cccrrrrrrc} 
        \hline\noalign{\smallskip}
SDSS NAME & $z_{\rm em}^{\rm a}$ & $z_{\rm abs}^{\rm b}$ & \makecell[c]{$F_{\rm cont1}^{\rm c}$} &  \makecell[c]{$F_{\rm cont2}^{\rm c}$} & \makecell[c]{$\Delta F_{\rm cont}$} & \makecell[c]{$\Delta EW_{\rm r} 48$} & \makecell[c]{$\Delta EW_{\rm r} 51$} & \makecell[c]{Index$^{\rm d}$} & \makecell[c]{Note$^{\rm b}$} \\
        \hline
005157.24+000354.7	&	1.9609 	&	1.8681 	&$	31.14 	\pm	1.51 	$&	$	19.60 	\pm	0.78 	$&	$	-0.23 	\pm	0.03 	$&	$	0.79 	\pm	0.25 	$&	$	0.64 	\pm	0.50 	$&	--1	&	Emerged	\\
015017.70+002902.4	&	3.0013 	&	2.8344 	&$	6.86 	\pm	0.83 	$&	$	5.27 	\pm	0.38 	$&	$	-0.13 	\pm	0.07 	$&	$	0.74 	\pm	0.35 	$&	$	0.72 	\pm	0.31 	$&	--1	&	Emerged	\\
020629.33+004843.1	&	2.4988 	&	2.3624 	&$	12.26 	\pm	1.05 	$&	$	5.88 	\pm	0.32 	$&	$	-0.35 	\pm	0.05 	$&	$	0.78 	\pm	0.39 	$&	$	0.76 	\pm	0.23 	$&	--1	&	Emerged	\\
024304.68+000005.4	&	2.0069 	&	1.9426 	&$	12.71 	\pm	1.65 	$&	$	19.80 	\pm	1.06 	$&	$	0.22 	\pm	0.07 	$&	$	-0.29 	\pm	0.05 	$&	$	-0.23 	\pm	0.07 	$&	--1	&	Weakened	\\
073232.79+435500.4	&	3.4618 	&	3.2569 	&$	7.48 	\pm	0.57 	$&	$	5.83 	\pm	0.33 	$&	$	-0.12 	\pm	0.05 	$&	$	0.80 	\pm	0.28 	$&	$	0.55 	\pm	0.20 	$&	--1	&	Strengthened	\\
073406.75+273355.6	&	1.9239 	&	1.8609 	&$	93.78 	\pm	2.68 	$&	$	83.53 	\pm	2.13 	$&	$	-0.06 	\pm	0.02 	$&	$	0.21 	\pm	0.05 	$&	$	0.34 	\pm	0.05 	$&	--1	&	Strengthened	\\
080006.59+265054.7	&	2.3438 	&	2.3033 	&$	10.23 	\pm	0.92 	$&	$	5.11 	\pm	0.48 	$&	$	-0.33 	\pm	0.06 	$&	$	0.67 	\pm	0.24 	$&	$	0.71 	\pm	0.38 	$&	--1	&	Emerged	\\
080609.24+141146.4	&	2.2877 	&	2.0062 	&$	19.14 	\pm	0.98 	$&	$	16.97 	\pm	0.55 	$&	$	-0.06 	\pm	0.03 	$&	$	-0.93 	\pm	0.19 	$&	$	-0.81 	\pm	0.18 	$&	1	&	Disappeared	\\
080906.88+172955.1	&	2.9770 	&	2.8828 	&$	14.23 	\pm	1.05 	$&	$	11.41 	\pm	0.43 	$&	$	-0.11 	\pm	0.04 	$&	$	0.71 	\pm	0.19 	$&	$	0.85 	\pm	0.88 	$&	--1	&	Emerged	\\
080906.88+172955.1	&	2.9770 	&	2.9384 	&$	14.23 	\pm	1.05 	$&	$	11.41 	\pm	0.43 	$&	$	-0.11 	\pm	0.04 	$&	$	0.63 	\pm	0.13 	$&	$	0.59 	\pm	0.11 	$&	--1	&	Strengthened	\\
081655.49+455633.7	&	2.7168 	&	2.5745 	&$	7.01 	\pm	0.78 	$&	$	7.99 	\pm	0.24 	$&	$	0.07 	\pm	0.06 	$&	$	0.68 	\pm	0.32 	$&	$	0.66 	\pm	0.25 	$&	1	&	Emerged	\\
081929.59+232237.4	&	1.8467 	&	1.7258 	&$	34.28 	\pm	1.39 	$&	$	22.85 	\pm	1.01 	$&	$	-0.20 	\pm	0.03 	$&	$	-0.49 	\pm	0.14 	$&	$	-0.52 	\pm	0.15 	$&	1	&	Weakened	\\
082751.78+132107.2	&	1.8289 	&	1.8019 	&$	42.91 	\pm	1.91 	$&	$	35.98 	\pm	1.18 	$&	$	-0.09 	\pm	0.03 	$&	$	0.17 	\pm	0.04 	$&	$	0.18 	\pm	0.05 	$&	--1	&	Strengthened	\\
091621.46+010015.4	&	2.2255 	&	2.1264 	&$	12.72 	\pm	0.84 	$&	$	9.15 	\pm	0.49 	$&	$	-0.16 	\pm	0.04 	$&	$	0.78 	\pm	0.30 	$&	$	0.70 	\pm	0.20 	$&	--1	&	Emerged	\\
091621.46+010015.4	&	2.2255 	&	2.1629 	&$	12.72 	\pm	0.84 	$&	$	9.15 	\pm	0.49 	$&	$	-0.16 	\pm	0.04 	$&	$	0.36 	\pm	0.05 	$&	$	0.26 	\pm	0.04 	$&	--1	&	Strengthened	\\
091621.46+010015.4	&	2.2255 	&	2.1741 	&$	12.72 	\pm	0.84 	$&	$	9.15 	\pm	0.49 	$&	$	-0.16 	\pm	0.04 	$&	$	0.40 	\pm	0.07 	$&	$	0.46 	\pm	0.11 	$&	--1	&	Strengthened	\\
095254.10+021932.8	&	2.1526 	&	2.0056 	&$	19.11 	\pm	1.15 	$&	$	21.37 	\pm	0.71 	$&	$	0.06 	\pm	0.03 	$&	$	0.67 	\pm	0.32 	$&	$	0.81 	\pm	0.50 	$&	1	&	Emerged	\\
100716.69+030438.7	&	2.1241 	&	1.9129 	&$	25.40 	\pm	1.20 	$&	$	16.00 	\pm	0.73 	$&	$	-0.23 	\pm	0.03 	$&	$	0.23 	\pm	0.05 	$&	$	0.35 	\pm	0.05 	$&	--1	&	Strengthened	\\
100716.69+030438.7	&	2.1241 	&	1.9426 	&$	25.40 	\pm	1.20 	$&	$	16.00 	\pm	0.73 	$&	$	-0.23 	\pm	0.03 	$&	$	0.86 	\pm	0.15 	$&	$	0.83 	\pm	0.27 	$&	--1	&	Emerged	\\
103115.69+374849.5	&	2.2590 	&	2.2100 	&$	8.24 	\pm	0.94 	$&	$	9.98 	\pm	0.51 	$&	$	0.10 	\pm	0.06 	$&	$	-0.78 	\pm	0.16 	$&	$	-0.75 	\pm	0.29 	$&	--1	&	Disappeared	\\
103115.69+374849.5	&	2.2590 	&	2.2253 	&$	8.24 	\pm	0.94 	$&	$	9.98 	\pm	0.51 	$&	$	0.10 	\pm	0.06 	$&	$	-0.77 	\pm	0.25 	$&	$	-0.85 	\pm	0.17 	$&	--1	&	Disappeared	\\
103842.14+350906.9	&	2.2049 	&	2.1563 	&$	13.52 	\pm	1.25 	$&	$	7.97 	\pm	0.48 	$&	$	-0.26 	\pm	0.05 	$&	$	0.35 	\pm	0.08 	$&	$	0.34 	\pm	0.08 	$&	--1	&	Strengthened	\\
104841.02+000042.8	&	2.0246 	&	1.9468 	&$	12.62 	\pm	1.02 	$&	$	13.87 	\pm	0.63 	$&	$	0.05 	\pm	0.05 	$&	$	-0.54 	\pm	0.08 	$&	$	-0.52 	\pm	0.11 	$&	--1	&	Weakened	\\
104923.94+012224.6	&	1.9454 	&	1.9087 	&$	37.98 	\pm	2.16 	$&	$	17.72 	\pm	0.79 	$&	$	-0.36 	\pm	0.03 	$&	$	0.74 	\pm	0.10 	$&	$	0.67 	\pm	0.13 	$&	--1	&	Strengthened	\\
105207.90+362219.4	&	2.3157 	&	2.2666 	&$	10.94 	\pm	1.13 	$&	$	14.78 	\pm	0.66 	$&	$	0.15 	\pm	0.06 	$&	$	0.32 	\pm	0.08 	$&	$	0.65 	\pm	0.15 	$&	1	&	Strengthened	\\
110726.04+385158.2	&	2.6603 	&	2.6134 	&$	13.70 	\pm	0.43 	$&	$	11.57 	\pm	0.55 	$&	$	-0.08 	\pm	0.03 	$&	$	-0.23 	\pm	0.05 	$&	$	-0.20 	\pm	0.07 	$&	1	&	Weakened	\\
115122.14+020426.3	&	2.4085 	&	2.3269 	&$	14.06 	\pm	1.02 	$&	$	7.88 	\pm	0.60 	$&	$	-0.28 	\pm	0.05 	$&	$	0.58 	\pm	0.10 	$&	$	0.52 	\pm	0.11 	$&	--1	&	Strengthened	\\
115122.14+020426.3	&	2.4085 	&	2.3742 	&$	14.06 	\pm	1.02 	$&	$	7.88 	\pm	0.60 	$&	$	-0.28 	\pm	0.05 	$&	$	0.18 	\pm	0.04 	$&	$	0.14 	\pm	0.04 	$&	--1	&	Strengthened	\\
120819.29+035559.4	&	2.0213 	&	1.9500 	&$	22.48 	\pm	1.88 	$&	$	12.93 	\pm	0.59 	$&	$	-0.27 	\pm	0.05 	$&	$	0.17 	\pm	0.04 	$&	$	0.26 	\pm	0.05 	$&	--1	&	Strengthened	\\
123720.85-011314.9	&	2.1620 	&	2.1283 	&$	10.22 	\pm	1.24 	$&	$	7.52 	\pm	0.46 	$&	$	-0.15 	\pm	0.07 	$&	$	0.78 	\pm	0.13 	$&	$	0.66 	\pm	0.17 	$&	--1	&	Strengthened	\\
124829.46+341231.3	&	2.2285 	&	2.0621 	&$	18.65 	\pm	1.24 	$&	$	8.88 	\pm	0.54 	$&	$	-0.35 	\pm	0.04 	$&	$	0.84 	\pm	0.20 	$&	$	0.87 	\pm	0.50 	$&	--1	&	Emerged	\\
124829.46+341231.3	&	2.2285 	&	2.0769 	&$	18.65 	\pm	1.24 	$&	$	8.88 	\pm	0.54 	$&	$	-0.35 	\pm	0.04 	$&	$	0.86 	\pm	0.30 	$&	$	0.83 	\pm	0.55 	$&	--1	&	Emerged	\\
125216.58+052737.7	&	1.9034 	&	1.8155 	&$	52.57 	\pm	2.29 	$&	$	88.82 	\pm	2.30 	$&	$	0.26 	\pm	0.02 	$&	$	-0.80 	\pm	0.22 	$&	$	-0.81 	\pm	0.12 	$&	--1	&	Disappeared	\\
125216.58+052737.7	&	1.9034 	&	1.8638 	&$	52.57 	\pm	2.29 	$&	$	88.82 	\pm	2.30 	$&	$	0.26 	\pm	0.02 	$&	$	-0.90 	\pm	0.06 	$&	$	-0.87 	\pm	0.18 	$&	--1	&	Disappeared	\\
125216.58+052737.7	&	1.9034 	&	1.8831 	&$	52.57 	\pm	2.29 	$&	$	88.82 	\pm	2.30 	$&	$	0.26 	\pm	0.02 	$&	$	-0.57 	\pm	0.03 	$&	$	-0.53 	\pm	0.05 	$&	--1	&	Weakened	\\
125216.58+052737.7	&	1.9034 	&	1.8946 	&$	52.57 	\pm	2.29 	$&	$	88.82 	\pm	2.30 	$&	$	0.26 	\pm	0.02 	$&	$	-0.75 	\pm	0.05 	$&	$	-0.76 	\pm	0.10 	$&	--1	&	Weakened	\\
132333.03+004750.2	&	1.7785 	&	1.7701 	&$	29.60 	\pm	2.69 	$&	$	34.12 	\pm	1.48 	$&	$	0.07 	\pm	0.05 	$&	$	-0.02 	\pm	0.04 	$&	$	-0.01 	\pm	0.04 	$&	--1	&	Weakened	\\
134544.55+002810.7	&	2.4641 	&	2.3514 	&$	11.21 	\pm	1.10 	$&	$	14.86 	\pm	0.54 	$&	$	0.14 	\pm	0.05 	$&	$	-0.87 	\pm	0.11 	$&	$	-0.89 	\pm	0.11 	$&	--1	&	Disappeared	\\
134544.55+002810.7	&	2.4641 	&	2.3686 	&$	11.21 	\pm	1.10 	$&	$	14.86 	\pm	0.54 	$&	$	0.14 	\pm	0.05 	$&	$	-0.93 	\pm	0.08 	$&	$	-0.90 	\pm	0.11 	$&	--1	&	Disappeared	\\
134544.55+002810.7	&	2.4641 	&	2.3964 	&$	11.21 	\pm	1.10 	$&	$	14.86 	\pm	0.54 	$&	$	0.14 	\pm	0.05 	$&	$	-0.40 	\pm	0.05 	$&	$	-0.54 	\pm	0.06 	$&	--1	&	Weakened	\\
140815.58+060023.3	&	2.5830 	&	2.5519 	&$	7.28 	\pm	0.80 	$&	$	4.80 	\pm	0.38 	$&	$	-0.20 	\pm	0.07 	$&	$	0.86 	\pm	0.75 	$&	$	0.85 	\pm	0.31 	$&	--1	&	Emerged	\\
150033.53+003353.6	&	2.4360 	&	2.1849 	&$	13.23 	\pm	1.04 	$&	$	18.97 	\pm	0.63 	$&	$	0.18 	\pm	0.04 	$&	$	0.62 	\pm	0.41 	$&	$	0.68 	\pm	0.25 	$&	1	&	Emerged	\\
160445.92+335759.0	&	1.8776 	&	1.7709 	&$	11.05 	\pm	1.41 	$&	$	9.34 	\pm	0.62 	$&	$	-0.08 	\pm	0.07 	$&	$	0.69 	\pm	0.39 	$&	$	0.75 	\pm	0.25 	$&	--1	&	Emerged	\\
160613.99+314143.4	&	2.0569 	&	2.0240 	&$	11.22 	\pm	1.21 	$&	$	10.72 	\pm	0.64 	$&	$	-0.02 	\pm	0.06 	$&	$	-0.35 	\pm	0.07 	$&	$	-0.41 	\pm	0.08 	$&	0	&	Weakened	\\
161336.81+054701.7	&	2.4855 	&	2.4152 	&$	9.03 	\pm	0.83 	$&	$	8.55 	\pm	0.46 	$&	$	-0.03 	\pm	0.05 	$&	$	0.66 	\pm	0.28 	$&	$	0.68 	\pm	0.27 	$&	--1	&	Emerged	\\
161511.35+314728.3	&	2.0981 	&	1.9157 	&$	41.00 	\pm	1.80 	$&	$	34.78 	\pm	1.11 	$&	$	-0.08 	\pm	0.03 	$&	$	-0.16 	\pm	0.04 	$&	$	-0.25 	\pm	0.05 	$&	1	&	Weakened	\\
162701.94+313549.2	&	2.3263 	&	2.2785 	&$	65.53 	\pm	2.40 	$&	$	62.42 	\pm	1.46 	$&	$	-0.02 	\pm	0.02 	$&	$	0.13 	\pm	0.03 	$&	$	0.25 	\pm	0.03 	$&	0	&	Strengthened	\\
162935.68+321009.5	&	2.0364 	&	1.9345 	&$	12.72 	\pm	1.44 	$&	$	9.55 	\pm	0.65 	$&	$	-0.14 	\pm	0.07 	$&	$	0.43 	\pm	0.13 	$&	$	0.35 	\pm	0.11 	$&	--1	&	Strengthened	\\
212943.25+003005.6	&	2.6802 	&	2.5726 	&$	8.11 	\pm	0.91 	$&	$	6.52 	\pm	0.36 	$&	$	-0.11 	\pm	0.06 	$&	$	0.77 	\pm	0.39 	$&	$	0.72 	\pm	0.18 	$&	--1	&	Emerged	\\
213648.17-001546.6	&	2.1736 	&	1.8372 	&$	8.35 	\pm	1.03 	$&	$	14.62 	\pm	0.57 	$&	$	0.27 	\pm	0.06 	$&	$	-0.89 	\pm	0.12 	$&	$	-0.86 	\pm	0.26 	$&	--1	&	Disappeared	\\
222157.97-010331.0	&	2.6744 	&	2.5459 	&$	15.66 	\pm	0.74 	$&	$	16.86 	\pm	0.51 	$&	$	0.04 	\pm	0.03 	$&	$	-0.31 	\pm	0.04 	$&	$	-0.34 	\pm	0.05 	$&	--1	&	Weakened	\\
230034.04-004901.5	&	2.2125 	&	2.1422 	&$	8.07 	\pm	0.95 	$&	$	13.67 	\pm	0.60 	$&	$	0.26 	\pm	0.06 	$&	$	-0.62 	\pm	0.07 	$&	$	-0.28 	\pm	0.07 	$&	--1	&	Weakened	\\
        \hline
    \end{tabular}
 \begin{tablenotes}
  \footnotesize
  \item \emph{Notes.}$^{\rm a}$Following Chen et al. (2015), the value of $z_{\rm em}$ is taken from Hewett \& Wild (2010).
  \item $^{\rm b}$The value of $z_{\rm abs}$ and the contents of "Note" are directly taken from Chen et al. (2015).
  \item $^{\rm c}$The values of $F_{\rm noise1}$ and $F_{\rm noise2}$ are measured in units of ${\rm 10^{-17}~erg~cm^{-2}~s^{-1}~\AA^{-1}}$.
  \item $^{\rm d}$The concordance index.

  \end{tablenotes}
\end{table*}
\subsection{Correlation analysis between C {\small IV} NALs and the continuum}
Based on the concordance index, we extracted three subsamples from
the whole sample. Subsample I includes the spectrum pairs with a
concordance index of $-1$. Subsample II includes those with a
concordance index of $+1$. Subsample III contains the spectrum pairs
of the emergence or disappearance of \ion{C}{iv} NALs included in
Subsample I. In the following, we analyse the
correlation between the variations of \ion{C}{iv} absorption-line
EWs and those of the continuum flux for the whole sample and the
above three subsamples, respectively.

Following Chen et al. (2015), we fitted a pseudo-continuum for each
spectrum using a combination of cubic spline functions and Gaussian
functions. Then we evaluated the pseudo-continuum fitting flux
variability amplitude between two observations for each quasar using
\begin{equation}
    \Delta F_{\rm cont}=\frac{F_{\rm cont2}-F_{\rm cont1}}{\textbar F_{\rm cont2}\textbar +\textbar F_{\rm cont1}\textbar},
    \label{eq:quadratic 1}
\end{equation}
where $F_{\rm cont1}$ and $F_{\rm cont2}$ denote the flux of the
pseudo-continuum at $\sim$ 1450 {\rm \AA} in the quasar rest-frame
for the earlier and later epochs, respectively. The propagation of
the error for $\Delta F_{\rm cont}$ can be evaluated using
\begin{equation}
\sigma_{\Delta F_{\rm cont}}=\sqrt{({\frac{\partial \Delta F_{\rm cont}}{\partial F_{\rm cont1}})}^2{F_{\rm noise1}}^2+{(\frac{\partial \Delta F_{\rm cont}}{\partial F_{\rm cont2}})}^2{F_{\rm noise2}}^2},
    \label{eq:quadratic 2}
\end{equation}
where $F_{\rm noise1}$ and $F_{\rm noise2}$ are the flux uncertainties
for the earlier and later SDSS epochs.

Then we calculated the variability amplitude of EWs  between the two
SDSS observations for the bluer and redder members ($\Delta EW_{\rm
r} 48$ and $\Delta EW_{\rm r} 51$) for each \ion{C}{iv}
$\lambda\lambda1548, 1551$ doublet using
\begin{equation}
\Delta EW_{\rm r} 48(51)=\frac{EW_{\rm r} 48(51)_{\rm 2}-EW_{\rm r} 48(51)_{\rm 1}}{\textbar EW_{\rm r} 48(51)_{\rm 2} \textbar+\textbar EW_{\rm r} 48(51)_{\rm 1} \textbar},
    \label{eq:quadratic 3}
\end{equation}
\begin{figure*}
    \includegraphics[width=\columnwidth]{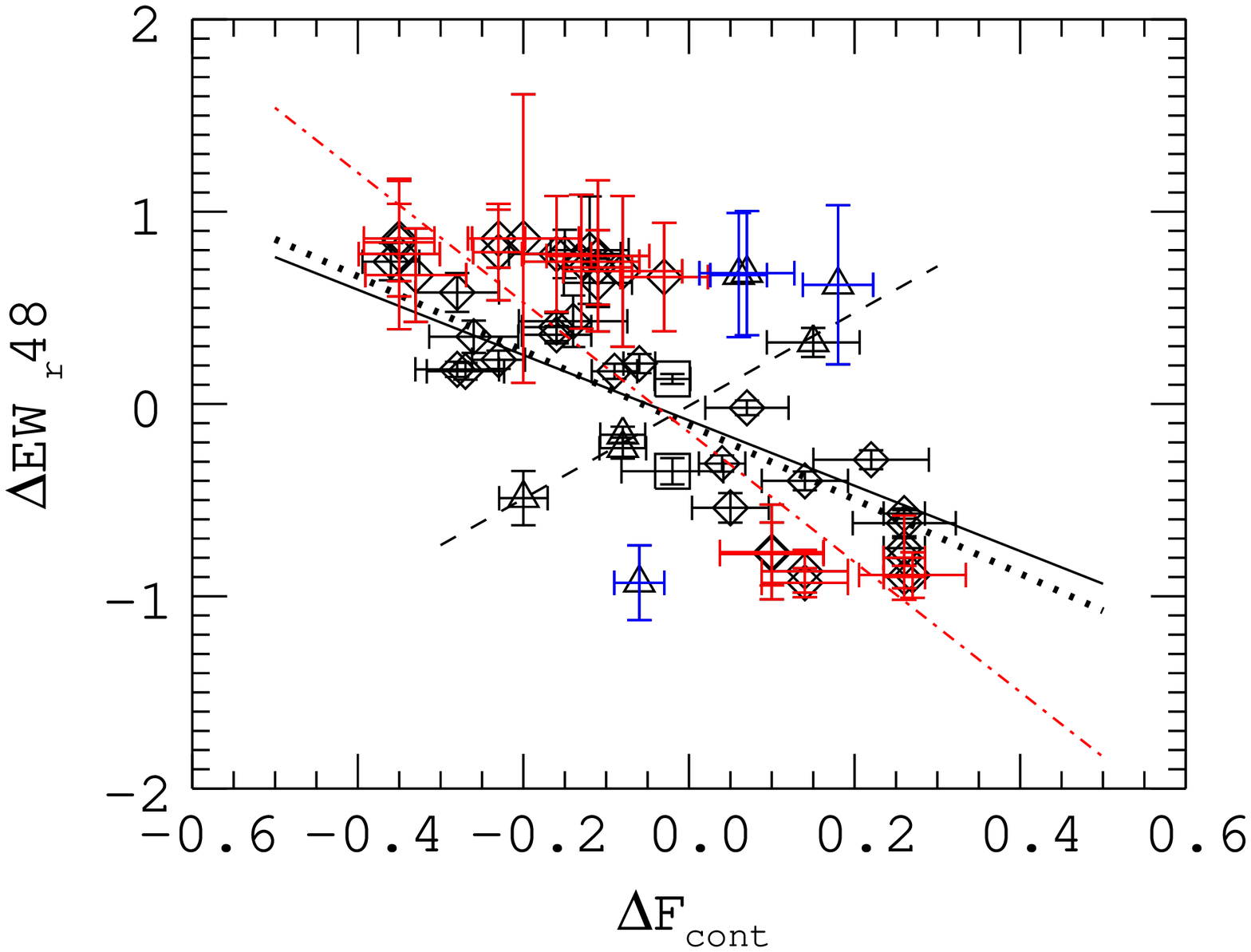}
    \includegraphics[width=\columnwidth]{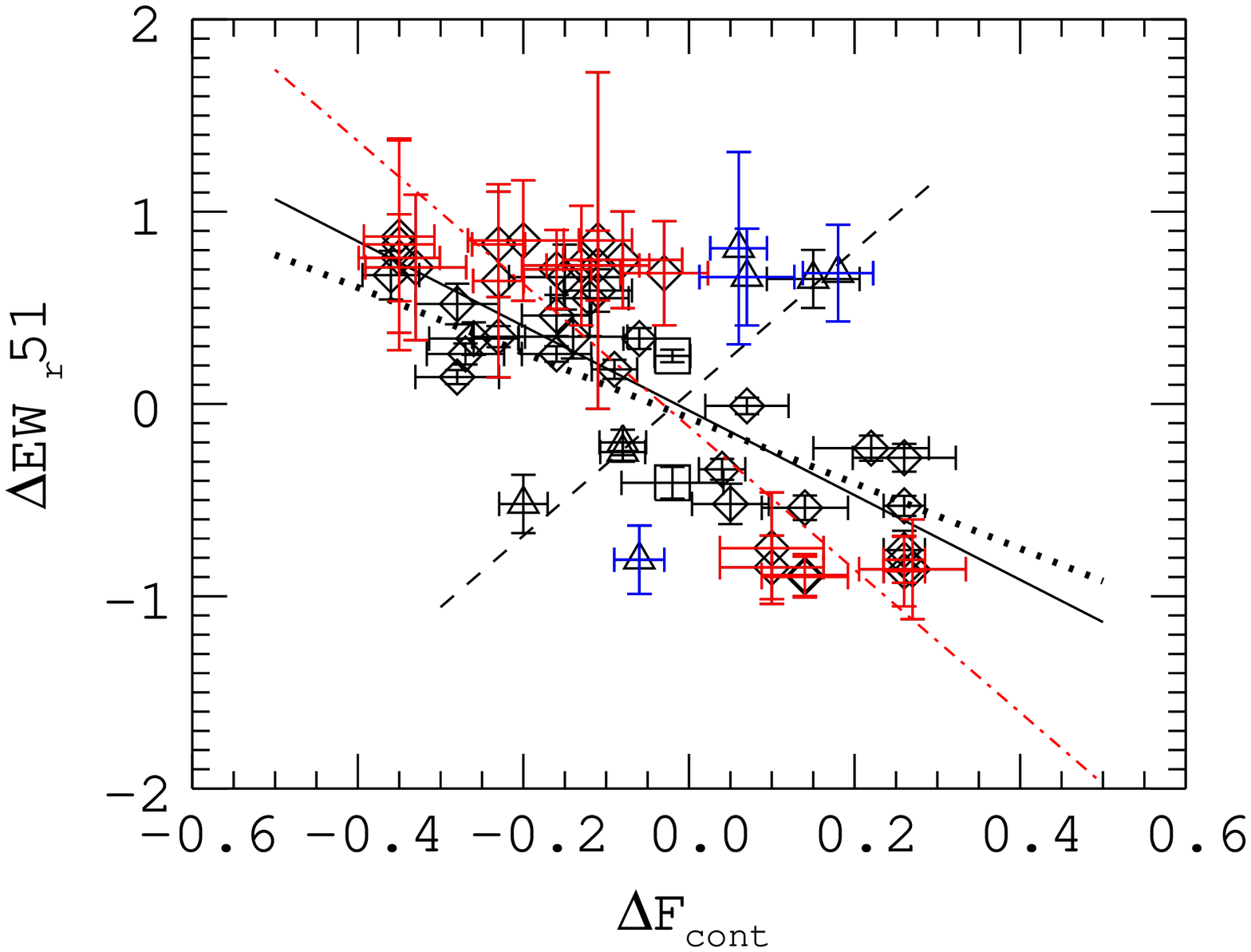}
 \caption{Plots of the variability amplitudes of narrow C {\scriptsize IV} absorption lines and the pseudo-continuum (left panel: C {\scriptsize IV} ${\rm \lambda}1548$ versus the pseudo-continuum;
  right panel: C {\scriptsize IV} ${\rm \lambda}1551$ versus the pseudo-continuum). The black rhombuses, black triangles and black squares represent the data with concordance indexes of $-1$, $+1$ and $0$, respectively. The black solid line, black dotted line, black dashed line and red dot-dashed line are the best linear fits for the whole sample, Subsample I, Subsample II and Subsample III, respectively. The errors are shown by the error bars. The data with red error bars are Subsample III. The data with blue error bars are the spectrum pairs of  emerging or disappearing absorption lines in Subsample II.}
   \label{fig:2}
\end{figure*}
\begin{table*}
    \centering
    \caption{Correlated parameters of $\Delta EW_{\rm r} 48$ ($\Delta EW_{\rm r} 51$) vs. $\Delta F_{\rm cont}$ for the whole sample and each subsample. The data in brackets denote the results of $\Delta EW_{\rm r} 51$ vs. $\Delta F_{\rm cont}$.}
    \label{tb:2}
    \begin{tabular}{ccccc} 
        \hline\noalign{\smallskip}
      &The whole sample &Subsample I&Subsample II &Subsample III\\
        \hline
$\emph N$$^{\rm a}$        &    52  &   42  &   8   &   20\\
$\emph r$$^{\rm b}$&    --0.699(--0.661)    &   --0.859(--0.848)    &   0.778(0.850)    &   --0.901(--0.893)\\
$\emph P$$^{\rm c}$&    \textless 0.001(\textless 0.001)    &   \textless 0.001(\textless 0.001)&\textless 0.05(\textless 0.01) &   \textless 0.001(\textless 0.001 )   \\
$\emph k$$^{\rm d}$ &   $-1.699\pm0.067$($-2.200\pm0.028$) &    $-1.933\pm0.067$($-1.697\pm0.068$)  &   $2.417\pm0.325$($3.714\pm0.485$)    &   $-3.376\pm0.177$($-3.715\pm0.247$)\\
$\emph b$$^{\rm e}$&    $-0.086\pm0.012$($-0.034\pm0.005$)  &   $-0.109\pm0.011$($-0.073\pm0.013$)  &   $-0.009\pm0.033$($0.058\pm0.047$)   &   $-0.147\pm0.039$($-0.119\pm0.048$)  \\
        \hline
    \end{tabular}
 \begin{tablenotes}
  \footnotesize
  \item[a] \emph{Notes.}$^{\rm a}$ The number of absorption systems in the samples.
  \item[b] $^{\rm b}$ The linear Pearson correlation coefficient.
  \item[c] $^{\rm c}$ The significance level of the Pearson correlation coefficient.
  \item[d] $^{\rm d}$ The slope of the best linear fit.
  \item[e] $^{\rm e}$ The intercept of the best linear fit.
  \end{tablenotes}
\end{table*}
where the EWs for \ion{C}{iv} $\rm \lambda 1548$ and \ion{C}{iv}
$\rm \lambda 1551$ were taken directly from Chen et al. (2015). We
also calculated their corresponding errors ($\sigma_{\Delta EW_{\rm
r} 48}$ and $\sigma_{\Delta EW_{\rm r} 51}$) with the same method as used in
calculating $\sigma_{\Delta F_{\rm cont}}$, respectively. The
parameters and  estimated values are listed in Table~\ref{tb:1}.

Here, we present plots of $\Delta EW_{\rm r} 48$ and $\Delta EW_{\rm
r} 51$ versus $\Delta F_{\rm cont}$ for all objects
(Fig.~\ref{fig:2}). Linear functions are adopted to fit the data
for the whole sample and the three subsamples, respectively. The
statistical parameters are listed in Table~\ref{tb:2}.

\section{Discussion}
\subsection{Properties of the variable C {\small IV} $\lambda\lambda1548, 1551$ absorption doublets}
We believe that the variable absorption systems in our sample are intrinsic to the corresponding quasars. On the one hand, the
result of the time variability analysis in this paper, namely the fact the variations of
narrow \ion{C}{iv} absorption lines show a significantly correlative
with the continuum variations, strongly supports the above suggestion. On the
other hand, the time variability of intervening cloud structures would
not be expected on such short time-scales because of their large
sizes and low densities (e.g., Hamann et al. 1995).

Based on the time variability analysis of 12 intrinsic NALs
and seven mini-BALs within the timescale of 1--3.5 yr in the quasar rest-frame, Misawa et al. (2014) found that only the latter show
significant variability. These authors suggested that changes in both the
ionizing continuum and the partial coverage can lead to the
variability of mini-BALs, and that the latter mechanism is the main driver
of NAL variability. However, our statistical analysis shows a
significant anti-correlation between the variation of narrow C
{\footnotesize IV} absorption lines and that of the pseudo-continuum flux
(see Fig.~\ref{fig:2}), implying that the NAL variability is
caused primarily by fluctuations in the ionizing continuum.

More specifically, we find that our sample of variable absorption lines
can be  divided mainly into two subsamples according to the
concordance index. Subsample I ($\thicksim$81 per cent) shows an
anti-correlation between the variability of NALs and the continuum
while Subsample II ($\thicksim$15 per cent) shows a positive
correlation. This phenomenon could be explained by the different
ionization states of the absorbers (Wang et al. 2015; Lu et al. in
preparation).

\subsection{On the emergence or disappearance of C {\small IV} $\lambda\lambda1548, 1551$ absorption doublets}
In our sample, there are 24 \ion{C}{iv} $\lambda\lambda1548, 1551$
absorption doublets that emerged in or disappeared from the BOSS spectra
(Chen et al. 2015), of which 20 absorption systems have a concordance
index of $-1$ (which are classified as Subsample III).

It is evident from Fig.~\ref{fig:2}
that this population is bimodal. The points with positive values of
$\Delta EW_{\rm r}48(51)$ have values between 0.62 and 0.87, while
the points with negative values of $\Delta EW_{\rm r}48(51)$ have
values between --0.75 to --0.93. We find that the population
distributes a narrow-value range of $\Delta EW_{\rm r}48(51)$ near
--0.7 to --0.9 or 0.7 to 0.9. According to equation (\ref{eq:quadratic 3}), $\Delta EW_{\rm r}48(51)$ in Subsample III should be $+1$ or
$-1$. The
deviations arise because, even in these extreme cases
where no absorption lines are observed, considering the error of EWs
between SDSS I/II and BOSS spectra, Chen et al. (2015) still allow a
try of fit to the `imagined' absorption troughs. As a result, a value of EWs is obtained that is slightly greater than zero instead of being
zero. Therefore, we find that the values of $\Delta EW_{\rm r}48(51)$ in
Subsample III are slightly lower than $+1$ or slightly greater than
$-1$. The observed bimodal population is expected, because while one
group is associated with the case of emergence, the other group
corresponds to the case of disappearance (note that the two groups
represent extreme cases of the corresponding variations).

In addition, it can be seen that the data points for
Subsample III span all values of  $\Delta F_{\rm cont}$. This
suggests that the emergence as well as the disappearance of NALs do not
always require that the continuum varies considerably, but, 
instead, emergence or disappearance could occur for even small variations. This implies that some absorption gas is very sensitive to
continuum variations, which was also suggested for the case of
BALs (Filiz Ak et al. 2013; Wang et al. 2015). We therefore suggest
that there might be two kinds of absorption gas: one that is very
sensitive to the continuum variations (which is expected to yield
data such as that in Subsample III), and aonther that is not so
sensitive (which is expected to produce the type of data other than
that in Subsample III). In Fig.~\ref{fig:2}, it is obvious that the
best linear fitting for Subsample III provides the steepest slope
for all points in the whole sample (this is to be expected, because data in
Subsample III reflect extreme cases of variation). Although some
cases of the emergence or disappearance of NALs might be caused by the bulk
motion of the absorption gas (Chen et al. 2013), the significant
anti-correlation between $\Delta EW_{\rm r}48(51)$ and $\Delta
F_{\rm cont}$ for Subsample III (see Fig.~\ref{fig:2}) strongly implies that
many extreme variations of NALs must also be
 due to the fluctuations
of the ionizing continuum.

It is clear that separate linear fits for the two extreme
groups (corresponding to the emergence or disappearance cases) in
Subsample III will produce very weak correlation between $\Delta
EW_{\rm r}48(51)$ and $\Delta F_{\rm cont}$. This is to be expected, because
for one group the value of $\Delta EW_{\rm r}48(51)$ is
expected to be $+1$, while it is expected to be $-1$ for the other
group. When ignoring the data of both of groups, there is still a significant anti-correlation between the two quantities
($r=-0.837$, $P<0.001$), suggesting that the existence of this
bimodal data of Subsample III strengthens the relationship of the
anti-correlation but the relationship itself is quite robust.

\subsection{Comparison of NAL and BAL/mini-BAL variability} 
We found that the variability of NALs is significantly correlated
with the continuum variability. However, as noted in the
Introduction, such a strong correlation between the variability of
BALs/mini-BALs and the continuum has not yet been found (although 
 coordinated variations between them have been detected; see Wang et
al. 2015). A plausible explanation for the different reactions
between BALs/mini-BALs and NALs is that BAL/mini-BAL outflows may
correspond to relatively larger clumpy/filament structures, whose UV
resonance doublets are usually blended and saturated, while the NALs
are caused by the smaller and lower-density portion of the outflow
wind (Murray et al. 1995; Proga, Stone \& Kallman 2000; Elvis 2000;
Ganguly et al. 2001a; Hall et al. 2007; Misawa et al. 2014), so that
NALs may be more sensitive than BALs regarding responses to the continuum
variability.

\section{Conclusion}

We have studied the correlation between the NAL variability and the
continuum variability using a sample of two-epoch SDSS spectra of 40
quasars with 52 variable narrow-absorption systems. We analysed
quantitatively the variability amplitude of EWs of \ion{C}{iv}
absorption lines and that of the continuum flux and found a
significant correlation between them. As hinted at by the statistical
results, we propose that the changes of the \ion{C}{iv}
$\lambda\lambda1548, 1551$ absorption doublets in our sample are
driven mainly  by the fluctuations of the ionizing continuum.

Our analysis suggests that the emergence and disappearance of
NALs does not require that the continuum must vary considerably.
Instead, these phenomena could occur for any level of variation
of the continuum. We suggest that there might be two kinds of
absorption gas: one that is very sensitive to the continuum
variations, and another that is not. Based on Fig.~\ref{fig:2}, we
believe that many cases of the emergence or disappearance of NALs are caused by fluctuations of the ionizing continuum as well.

\section*{Acknowledgements}

We thank the anonymous referee for helpful comments. This work was supported by the Science Research Projects of Guangxi Colleges and
Universities (No. KY2015YB289), and the Key Projects of Baise University (No.
2015KAN04).

Funding for SDSS-III was provided by the Alfred P. Sloan Foundation, the
Participating Institutions, the National Science Foundation, and the US
Department of Energy Office of Science. The SDSS-III web site is
http://www.sdss 3.org/.

SDSS-III is managed by the Astrophysical Research Consortium for the
Participating Institutions of the SDSS-III Collaboration, including the
University of Arizona, the Brazilian Participation Group, Brookhaven National
Laboratory, Carnegie Mellon University, University of Florida, the French
Participation Group, the German Participation Group, Harvard University, the
Instituto de Astrofisica de Canarias, the Michigan State/Notre Dame/JINA
Participation Group, Johns Hopkins University, Lawrence Berkeley National
Laboratory, Max Planck Institute for Astrophysics, Max Planck Institute for
Extraterrestrial Physics, New Mexico State University, New York University, Ohio
State University, Pennsylvania State University, University of Portsmouth,
Princeton University, the Spanish Participation Group, University of Tokyo,
University of Utah, Vanderbilt University, University of Virginia, University of
Washington, and Yale University.










\bsp    
\label{lastpage}
\end{document}